\documentclass[aps,pra,showpacs,groupedaddress,twocolumn,showkeys]{revtex4-2}
\usepackage{amsmath}
\usepackage{graphicx}
\usepackage{diagbox}
\usepackage{xcolor}
\usepackage[colorlinks=true, letterpaper=true, pdfstartview=FitV, linkcolor=blue, citecolor=blue, urlcolor=blue]{hyperref}

\begin{document}
	
\title{Parametrically driven pure-quartic solitons}

\author{Pengfei Li$^{1,2}$}
\email{lpf281888@gmail.com}
\author{Lijing Xing$^{1,2}$}
\author{Dongdong Wang$^{1,2}$}
\author{Dumitru Mihalache$^{1,2}$}
\author{David Laroze$^{1,2}$}
\author{Boris A. Malomed$^{4,5}$}

\affiliation{$^{1}$Department of Physics, Taiyuan Normal University, Jinzhong, Shanxi 030619, China}
\affiliation{$^{2}$Shanxi Key Laboratory for Intelligent Optimization Computing and Blockchain Technology, Taiyuan Normal University, Jinzhong, Shanxi 030619, China}
\affiliation{$^{3}$Horia Hulubei National Institute of Physics and Nuclear Engineering, Magurele, Bucharest RO-077125, Romania}
\affiliation{$^{4}$Instituto de Alta Investigaci\'{o}n, Universidad de Tarapac\'{a}, Casilla 7D, Arica, Chile}
\affiliation{$^{5}$Department of Physical Electronics, School of Electrical Engineering, Faculty of Engineering, and Center for Light-Matter Interaction, Tel Aviv University, Tel Aviv 69978, Israel}

\begin{abstract}
Parametrically driven solitons are self-trapped modes in various physical settings, including optics, magnetics, etc. So far, the analysis was focused on the existence, stability, and dynamics of such solitons in systems including the second-order group-velocity dispersion (GVD), linear loss, parametric gain, and cubic nonlinearity. Here, we report the existence of quiescent parametrically driven pure-quartic solitons (PDPQSs) in the full system, and moving PDPQSs in the absence of losses. A systematic analysis reveals stability domains for the solitons in the system’s parameter space. Evolution of unstable states is explored too, and it is demonstrated that collisions between traveling stable PDPQSs are elastic.
\end{abstract}

\maketitle

\section{Introduction}

Systems governed by different physical laws may be described by the same
effective equations, thereby establishing connections between different
areas. A prominent example is provided by the parametrically driven damped
nonlinear Schr\"{o}dinger equation (PDDNLSE), which accounts for pattern
formation phenomena in a range of physical settings, including surface
water-wave solitons \cite{Ref1,Ref2,Ref3}, light pulses in optical fibers
\cite{Ref4}, and solitons in magnetics \cite{Ref5}. In photonics, solitary
waves produced by PDDNLSE have been predicted in optical parametric
oscillators \cite{Ref6,Ref7,Ref8} and mode-locked lasers \cite{Ref9}. Much
interest has been drawn to solitons of this type maintained by
parametrically driven Kerr cavities \cite{Ref10,Ref11}, due to their
potential applications to optical communication \cite{Ref12}, high-precision
metrology \cite{Ref13}, parallel optical computing \cite{Ref14}, and
spectroscopy \cite{Ref15}.

A remarkable feature of the PDDNLSEs with the focusing \cite{Ref16,Ref17} or
defocusing \cite{Ref18} cubic self-phase-modulation (SPM) terms is the
existence of exact stationary soliton solutions. The PDDNLSEs also maintain
various traveling-soliton solutions \cite{Ref19,Ref19a,Ref20,Ref21}, some of which
are stable \cite{Ref21}. Furthermore, damped-driven solitons can form bound
states in the form of stationary and oscillatory complexes \cite{Ref22}. The
bound states, along with the traveling-soliton solutions in the conservative
limit, have been studied both analytically and numerically, revealing
sophisticated stability landscapes and complex dynamical regimes that
include oscillatory soliton pairs and moving localized structures \cite%
{Ref23,Ref24}.

So far, in studies of these systems, the focus has been on the balance
between the SPM nonlinearity and second-order dispersion. Recently, a new
class of \textit{pure-quartic solitons} (PQSs), has been identified,
arising from the balance between the fourth-order group-velocity dispersion
(GVD) and cubic SPM \cite{LPR7177}. PQSs have been predicted and experimentally observed in
photonic-crystal waveguides by means of precise GVD engineering \cite{Ref25}%
, using a mode-locked laser incorporating an intra-cavity spectral pulse
shaper to nullify the usual second-order GVD and make the quartic dispersive
term the leading one \cite{Ref26,Ref27}. In this vein, PQS solutions for
optical dissipative solitons in waveguides with the dominant quartic GVD
offer advantages for the realization in fiber lasers, such as flatter
spectra and the favorable energy scaling with respect to the pulse's
temporal width \cite{Ref28,Ref29,Ref30}. These properties suggest
possibilities for increasing the power of ultra-short solitons generated by
high-Q microring resonators \cite{Ref31}. In this context, dissipative PQSs,
produced by the Lugiato-Lefever equation, have been predicted in microrings
\cite{Ref32} and shown to stay robust under the action of the Raman
scattering \cite{Ref33,Ref34,Ref35} and pure high-order GVD \cite{Ref36},
where the gain, compensating the background loss, is provided by the ac drive.
While the dissipative PQSs have been widely studied in passively mode-locked lasers and directly driven Kerr cavities, the possibility of the existence of such solitons in parametrically driven systems was not considered. In this Letter, we demonstrate that the parametric drive, distinct from the mode-locking and direct driving, maintains parametrically-driven PQSs (PDPQSs) through a robust balance with the loss, GVD, and nonlinearity. In this context, we study the existence, stability, and dynamics of PDPQSs in the framework of the PDDNLSE, which is a relevant model of fiber cavities. Both quiescent and moving PDPQSs are identified, the latter ones existing below a critical velocity,  $V_{cr}$. One of the soliton species revealed by the analysis is stable in a broad parameter domain, while others are unstable. The moving PDPQSs exhibit elastic head-on and pursuit collisions.
\section{Model and Methods}

The starting point is the scaled PDDNLSE in the moving reference frame \cite%
{Ref19}, with the fourth-order dispersion (FOD) and focusing SPM:
\begin{equation}
i\frac{\partial \psi }{\partial t}-iV\frac{\partial \psi }{\partial \tau }%
+(i\gamma +\omega )\psi -\frac{1}{24}\frac{\partial ^{4}\Psi }{\partial \tau
^{4}}+2|\psi |^{2}\psi =h\psi ^{\ast },  \label{FODPDDNLSE}
\end{equation}%
where $\psi (t,\tau )$ is the complex field amplitude, $t$ and $\tau $ are
the slow and fast temporal variables, respectively. $V$ is the velocity of
moving solitons, $\gamma \geq 0$ is the loss coefficient, real $\omega $ and
$h>0$ are, respectively, the detuning and strength of the parametric drive,
and $\ast $ stands for the complex conjugate. 

The physical system modeled by Eq. (\ref{FODPDDNLSE}) can be implemented using a fiber cavity with an inner phase-sensitive amplifier \cite{Ref10} and a programmable spectral pulse shaper used to control the net cavity dispersion, cf. Ref. \cite{Ref27}. This setup can be used as a promising platform for the realization of advanced nonlinear pulse-shaping regimes, including the PDPQSs considered below.

We look for stationary PDPQS solutions of Eq. (\ref{FODPDDNLSE}) by solving
the respective ordinary differential equation,
\begin{equation}
-iV\frac{d\psi }{d\tau }+(i\gamma +\omega )\psi -\frac{1}{24}\frac{d^{4}\Psi
}{d\tau ^{4}}+2|\psi |^{2}\psi =h\psi ^{\ast },  \label{ODE-FODPDDNLSE}
\end{equation}%
under the zero boundary conditions, $|\psi |\rightarrow 0$ as $\tau
\rightarrow \infty $. Setting $\psi (\tau )\equiv \psi _{r}(\tau )+i\psi
_{i}(\tau )$, we solved Eq. (\ref{ODE-FODPDDNLSE}) by means of the
Newton-conjugate gradient method \cite{Ref37}. With an appropriate initial
guess, the method converges to numerical solutions through successive
iterations.

To test the stability of the PDPQS solutions, we computed eigenvalues of
small perturbations added to the PDPQS solutions, using the Fourier
collocation method \cite{Ref37}, and then verified the results by simulating
the perturbed evolution. The perturbed solution was introduced as $\psi
=\psi _{r}(\tau )+i\psi _{i}(\tau )+\left[ u(\tau )+iv(\tau )\right] \exp
(\delta t)$, where $u(\tau )$ and $v(\tau )$ are components of the eigenmode
corresponding to complex eigenvalue $\delta $. The substitution of this in
Eq. (\ref{FODPDDNLSE}) and linearization leads to the system of coupled
equations:
\begin{equation}
\delta u=+\frac{1}{24}\frac{d^{4}v}{d\tau ^{4}}-2(\psi _{r}^{2}+3\psi
_{i}^{2})v-(\omega +h)v+V\frac{du}{d\tau }-4\psi _{r}\psi _{i}u-\gamma u,
\label{Linearization_a}
\end{equation}%
\begin{equation}
\delta v=-\frac{1}{24}\frac{d^{4}u}{d\tau ^{4}}+2(3\psi _{r}^{2}+\psi
_{i}^{2})u+(\omega -h)u+V\frac{dv}{d\tau }+4\psi _{r}\psi _{i}v-\gamma v,
\label{Linearization_b}
\end{equation}%
which were solved by the Fourier collocation method. The solitons
are unstable if there are eigenvalues with $\mathrm{Re}(\delta )>0$.

\section{Numerical Results}

\begin{figure*}[ht]
	\centering
	\includegraphics[width=0.8\textwidth]{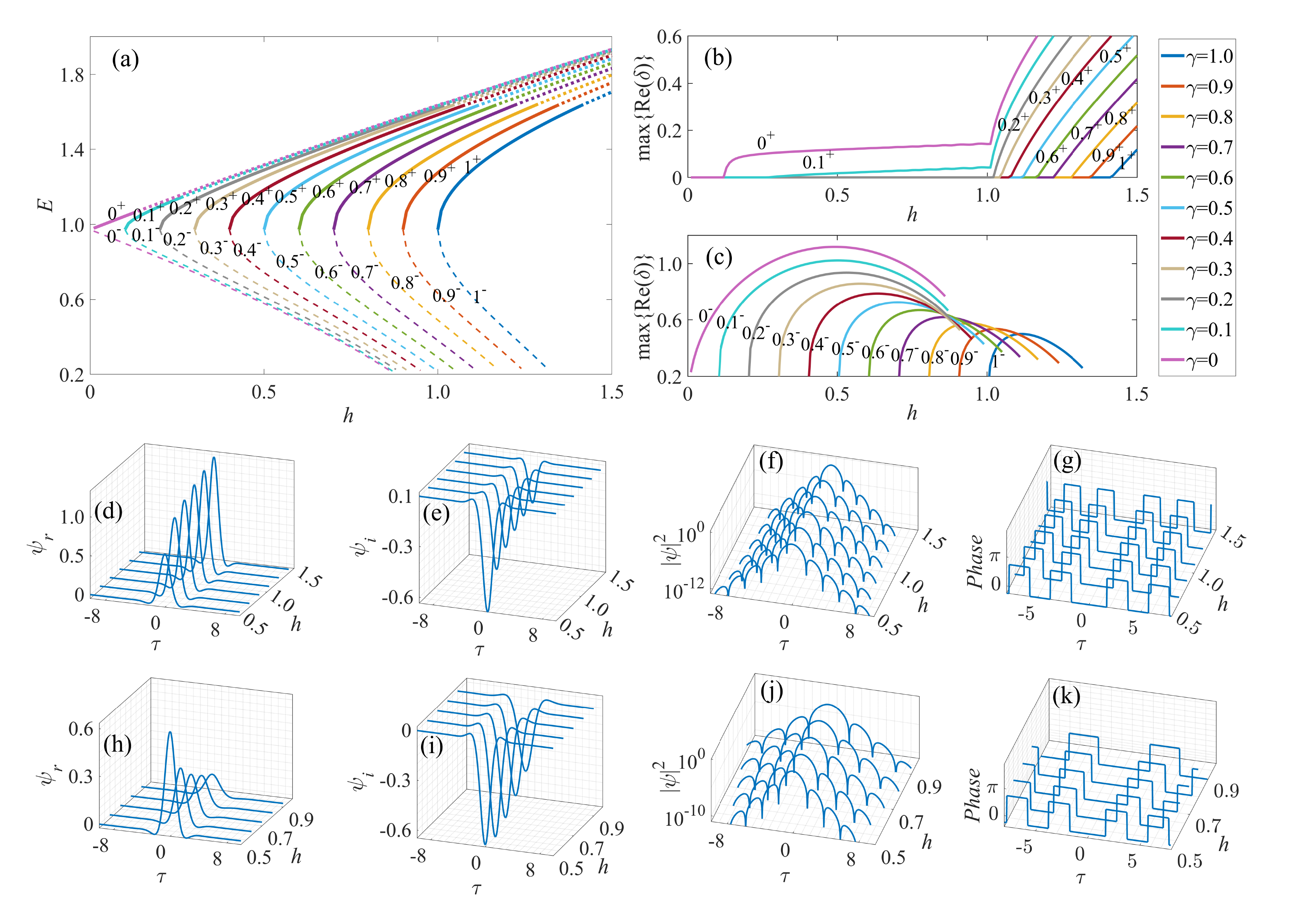}
	\caption{The existence and stability chart for the quiescent PDPQSs. (a) The
		bifurcation diagram for the conservative ($\protect\gamma =0$) and
		dissipative ($\protect\gamma >0$) systems is represented by dependence of
		the soliton's energy $E$ on the parametric-drive strength $h$. Thick and
		thin lines denote the upper ($\protect\psi _{+}$) and lower ($\protect\psi %
		_{-}$) branches, while solid and dashed or dotted lines denote stable and
		unstable families, respectively. Labels attached to the branches mark the
		corresponding values of $\protect\gamma $, with the superscripts $+$ and $-$
		indicating the solitons of the $\protect\psi _{+}$ and $\protect\psi _{-}$
		types, respectively. The maximum instability growth rates $\max \left\{
		\mathrm{Re}\left( \protect\delta \right) \right\} $ are plotted in panels
		(b) and (c) for the upper and lower branches, respectively. The bottom panels display the variation of profiles of the real and imaginary components (d and e, respectively), intensities on the logarithmic scale (f) and phase patterns (g) of the upper-branch solitons, with varying $h$ and fixed $\protect\gamma =0.5$ (these solitons are stable at $h<1.11$). (h)-(k) The same as in (d)-(g), but for unstable lower-branch solitons.}
	\label{fig1}
\end{figure*}

We first focus on the quiescent solitons of Eq. (\ref{ODE-FODPDDNLSE}) with $%
V=0$, fixing $\omega =-1$ by means of scaling. Our findings are summarized
in Fig. \ref{fig1}, where the stability results are also included. Two types
of quiescent PDPQSs, $\psi _{+}$ and $\psi _{-}$, are presented in Fig. \ref%
{fig1}(a), for different fixed values of loss constant $\gamma $ (including $%
\gamma =0$), by means of the upper ($\psi _{+}$) and lower ($\psi _{-}$)
branches of the dependence of the soliton's energy (alias the integral
norm), $E=\int_{-\infty }^{+\infty }|\psi |^{2}d\tau $, on strength $h$ of
the parametric drive. It is seen that both branches emerge through a
bifurcation of the saddle-node type precisely at $h=\gamma $, and exist at $%
h>\gamma $ (the parametric drive cannot balance the loss at $h<\gamma $ \cite%
{Ref5}). The upper branch $\psi _{+}$ of the quiescent PDPQSs is initially
stable, losing its stability at some critical value of $h$, as seen in Fig. %
\ref{fig1}(b). On the other hand, it was found that all the quiescent PDPQSs
belonging to the lower branches ($\psi _{-}$) in Fig. \ref{fig1}(c) are
completely unstable.

Typical examples of the temporal profiles of the quiescent PDPQSs belonging
to the upper branch ($\psi _{+}$), for a fixed loss coefficient $\gamma =0.5$%
, are plotted in Figs. \ref{fig1}(d,e,f). As seen in Fig. \ref{fig1}(f), the
profiles exhibit oscillatory tails, which is a hallmark of PQSs \cite{Ref25}%
. Phase patterns of the corresponding quiescent PDPQSs are displayed in Fig. %
\ref{fig1}(g), showing an essential difference from the usual parametrically
driven dissipative solitons, cf. Ref. \cite{Ref16}. Profiles of unstable
solitons, belonging to the lower branch $\psi _{-}$, and their phase
patterns are displayed in Figs. \ref{fig1}(h,i,j), and (k), respectively.

\begin{figure}[!ht]
	\centering
	\includegraphics[width=0.72\linewidth]{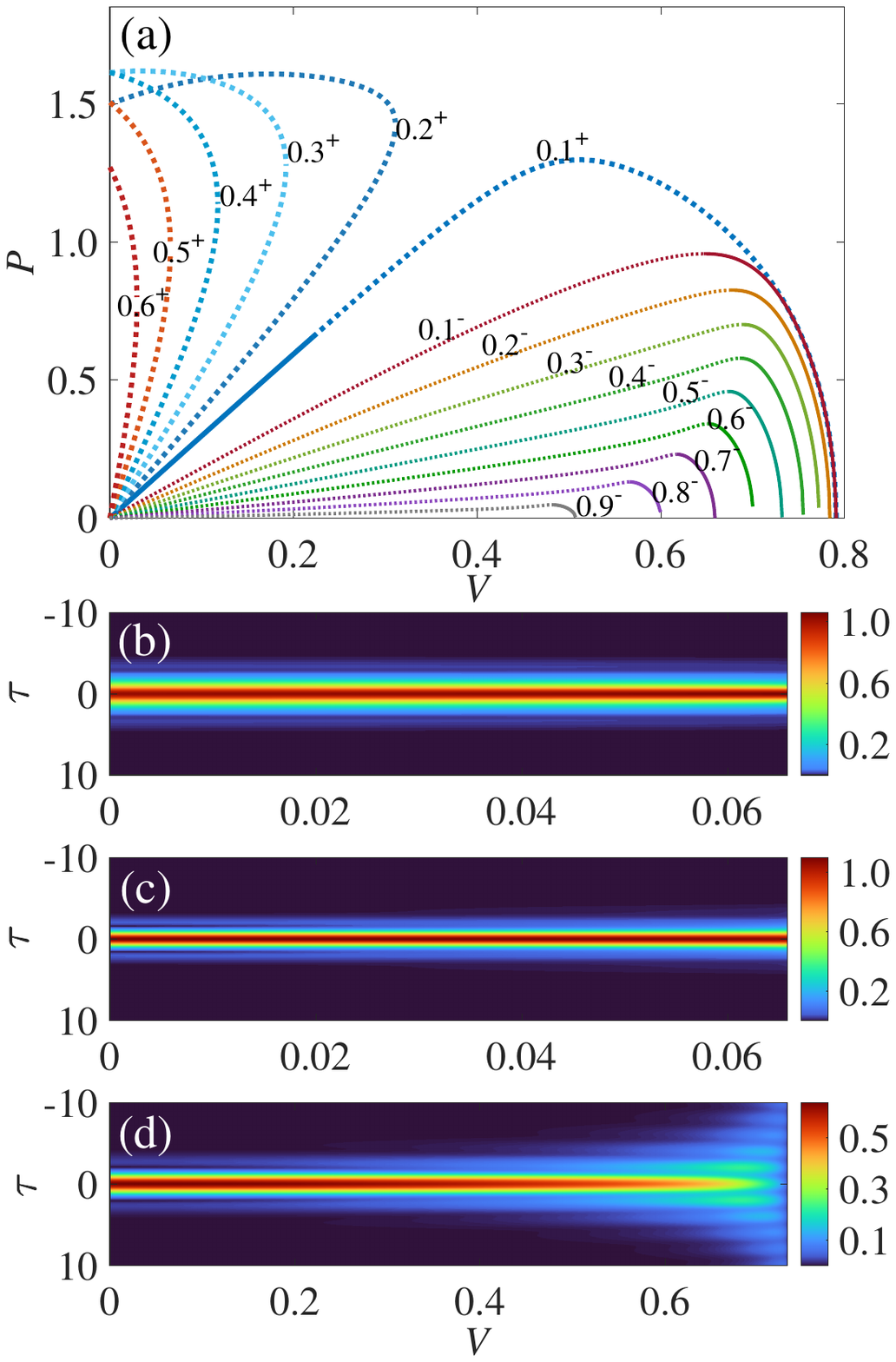}
	\caption{The existence and stability chart for the moving PDPQSs in the
		lossless case, $\protect\gamma =0$. (a) The momentum $P$ of the moving
		solitons [see Eq. (\protect\ref{eq:momentum})] vs. their velocity $V$. Solid
		and dotted lines designate stable and unstable solutions, respectively.
		Labels attached to the branches mark the corresponding values of $h$, with
		the superscripts $+$ and $-$ indicating the extension of the solitons of
		the $\protect\psi _{+}$ and $\protect\psi _{-}$ types, respectively, in
		terms of Fig. \protect\ref{fig1}. Panels (b) and (c) display the variation
		of profiles of $|\protect\psi _{+}|$ with $h=0.5$ following the increase of
		velocity $V$, along the upper and lower branches, respectively. (d) The same
		as in (b) and (c), but for $|\protect\psi _{-}|$ with $h=0.5$, the solitons
		being stable at $V>0.686$.}
	\label{fig2}
\end{figure}

Moving solitons can be used in all-optical data-processing systems and all-optical routing. To address solutions of Eq. (\ref{ODE-FODPDDNLSE}) for moving PDPQSs in the conservative system ($\gamma= 0$), we again fix $\omega =-1$ by means of scaling. As shown in Fig. \ref{fig2}, the moving PDPQSs exist up to a certain maximum value of the velocity, suffering delocalization beyond this point. The existence and stability of the moving solitons is presented in Fig. \ref{fig2}(a) by the dependence of the soliton’s momentum
\begin{equation}
	P=\frac{i}{2}\int_{-\infty }^{+\infty }\left( \frac{d\psi ^{\ast }}{d\tau }%
	\psi -\frac{d\psi }{d\tau }\psi ^{\ast }\right) d\tau ,  \label{eq:momentum}
\end{equation}%
on its velocity $V$. In the lossless system, $P$ is a conserved quantity,
along with the above-mentioned energy $E$.

Two types of moving PDPQSs are produced by Eq. (\ref{ODE-FODPDDNLSE}),
corresponding to the $\psi _{+}$ and $\psi _{-}$ soliton branches, in terms
of Fig. \ref{fig1}. For $h\geq 0.2$, the $\psi _{+}$ family of moving PDPQSs
in Fig. \ref{fig2} includes upper and lower branches, originating from $V=0$
and coalescing at a critical velocity $V=V_{\max }$, both branches being
completely unstable. Typical examples of the temporal profiles ($|\psi |$)
of the moving PDPQSs of the $\psi _{+}$ type, belonging to the upper and
lower branches at $h=0.5$, are plotted in Figs. \ref{fig2}(b) and (c),
respectively.

At $h<0.2$, only a single branch of the moving solitons of the $\psi _{+}$
type was found, extending continuously up to $V=V_{\mathrm{cr}}^{+}$, at
which point momentum $P$ vanishes. In this case, \emph{stable} moving PDPQSs
of the $\psi _{+}$ type are identified in the region $0\leq V\leq 0.221$.
For each value of $h$, the branch of moving PDPQSs of the $\psi _{-}$ type
also terminates at the corresponding point $V=V_{\mathrm{cr}}^{-}$. As $%
V\rightarrow V_{\mathrm{cr}}^{-}$, the moving PDPQSs of the $\psi _{-}$ type
develops oscillations in its tails, while the width of the resulting
oscillatory profile grows and the amplitude decreases, eventually
degenerating into $\psi \equiv 0$, as shown in Fig. \ref{fig2}(d). The
moving PDPQSs of the $\psi _{-}$ type are unstable for $dP/dV>0$ and stable
for $dP/dV<0$. This conclusion agrees with the well-known fact that the
point of $dP/dV=0$ is the stability border for traveling solitons \cite%
{Ref19}.

\begin{figure}[!ht]
	\centering
	\includegraphics[width=0.75\linewidth]{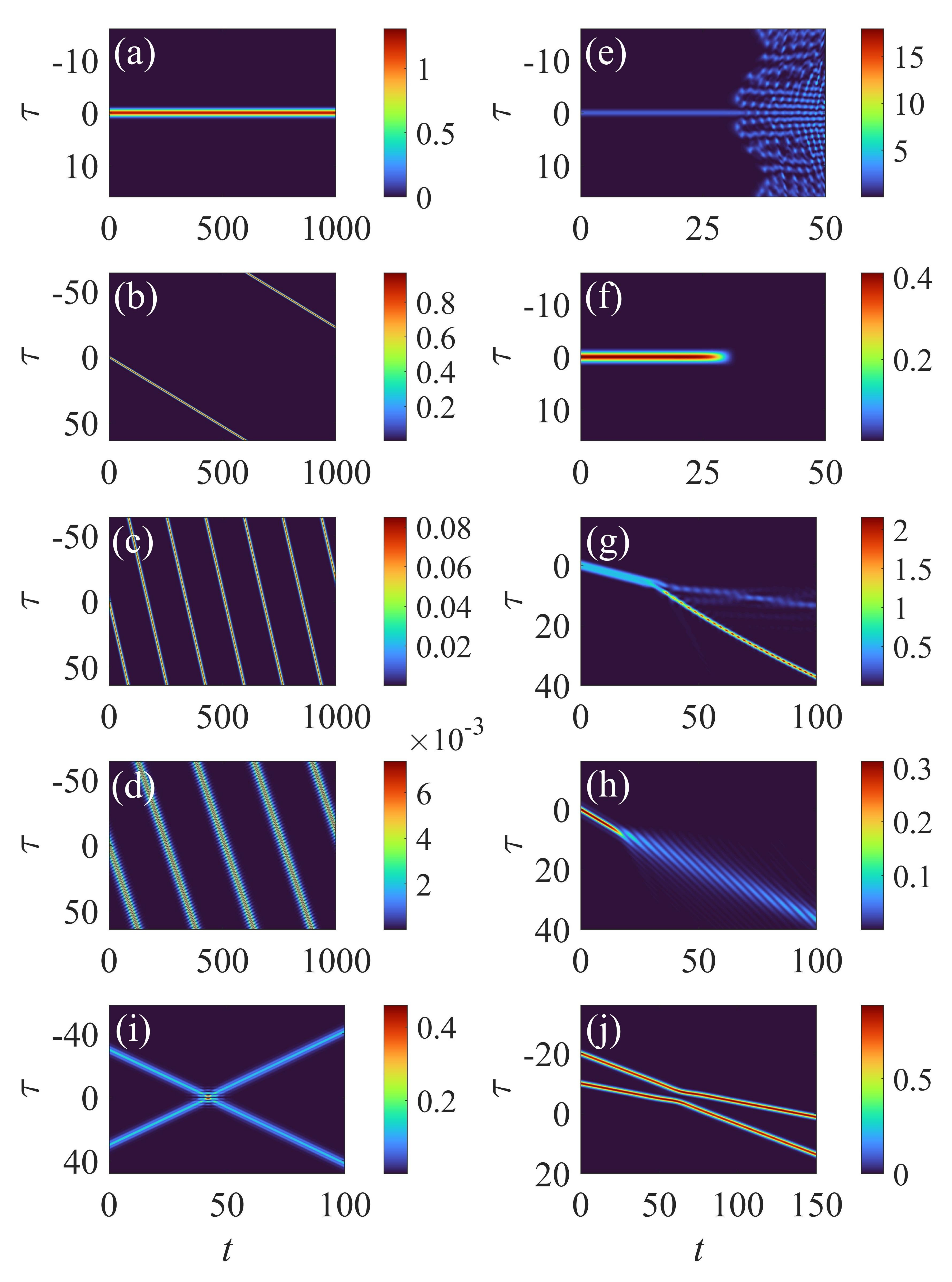}
	\caption{Evolution and interactions of PDPQSs. The perturbed dynamical
		evolutions: (a) quiescent PDPQS of $\protect\psi _{+}$ type for $\protect%
		\gamma =0.5$ with $h=0.7$, (b) moving PDPQSs of the $\protect\psi _{+}$ type
		traveling with velocity $V=0.1$ for $h=0.1$, the $\protect\psi _{-}$ type
		traveling with velocities $V=0.75$ for $h=0.1$ (c) and $V=0.5$ for $h=0.9$
		(d). The initial random perturbations in the simulations are taken with the $%
		5\%$ amplitude level. Evolutions of unstable quiescent PDPQSs: (e) $\protect%
		\psi _{+}$ type for $\protect\gamma =0.4$ with $h=1.45$, and (f) $\protect%
		\psi _{-}$ type for $\protect\gamma =0.5$ with $h=0.7$. Evolution of
		unstable moving PDPQSs: $\protect\psi _{+}$ type traveling with velocity $%
		V=0.2$ for $h=0.2$ (g), and $\protect\psi _{-}$ type with $V=0.465$ for $%
		h=0.5$ (h). The collision and pursuit interaction scenarios for a pair of
		stable traveling PDPQSs. (i) The collision between the solitons of the $%
		\protect\psi _{-}$ type, with $V_{1}=-V_{2}=0.7$ for $h=0.3$. (j) The
		pursuit between the solitons of the $\protect\psi _{+}$ type, with $V_{1}=0.2
		$ , $V_{2}=0.1$, for $h=0.1$.}
	\label{fig3}
\end{figure}

Finally, the dynamics and interactions (collisions) of the PDPQSs were
systematically simulated, in the framework of Eq. (\ref{FODPDDNLSE}), using
the Runge-Kutta method. Typical examples of the stable evolution of PDPQSs
are displayed in Figs. \ref{fig3}(a-d). The results confirm that quiescent
and moving PDPQSs with the stable spectrum of eigenvalues of small
perturbations indeed remain stable, at least, up to $t=1000$, under the
action of random initial perturbations with a relatively large amplitude of $%
5\%$. Representative examples, revealing the complex dynamics of unstable
solitons, are illustrated in Figs. \ref{fig3}(e-h). Specifically, unstable
quiescent PDPQS of the $\psi _{+}$ type suffer fragmentation in Fig. \ref%
{fig3}(e), while unstable quiescent PDPQSs of the $\psi _{-}$ type gradually
decay and eventually vanish, as shown in Fig. \ref{fig3}(f). Unstable moving
PDPQSs of the $\psi _{+}$ type, after passing a short distance, rapidly
break up, see Fig. \ref{fig3}(g). Unstable moving profiles of the $\psi _{-}$
type lose their stationary shape and develop intensive pulsation, as shown
in Fig. \ref{fig3}(h).

To study head-on collisions between stable moving PDPQSs in the conservative system ($\gamma= 0$), the simulations were initiated by the input in the form of a soliton pair with opposite velocities. As shown in Fig. \ref{fig3}(i), the solitons keep their original shapes after the head-on collision, with no apparent deformation or energy loss. We have also performed simulations for the pursuit collisions, launching two stable solitons with different velocities in the same direction (this case is not tantamount to the head-on collision, as the system is not a Galilean-invariant one). In this case as well, the solitons preserve their profiles in the post-collision state, see Fig. \ref{fig3}(j). As Eq. (\ref{FODPDDNLSE}) is far from any integrable model, the quasi-elasticity of the collisions is a nontrivial finding. These results also confirm that PDPQS in the conservative system are truly robust modes, as they readily survive collisions.

\section{Conclusion}
In conclusion, we have theoretically investigated the existence and properties of the new soliton species, PDPQSs (parametrically driven pure-quartic solitons), and identified parameter regions in which the quiescent ones are stable. These pulses arise from the double balance between the loss and parametric drive, as well as self-phase modulation and pure fourth-order dispersion. We have also studied the moving PDPQSs (in the absence of loss), and collisions between them (which turn out to be elastic). Unlike the usual parametrically driven solitons, PDPQSs feature oscillatory tails and an intrinsic phase structure. It should be stressed that, while the stable quiescent PDPQSs in the presence of the loss and stably moving ones in the absence of loss have been found in this work, the possible formation of stable traveling PDPQSs in the presence of the loss remains a subject for future research.

\begin{acknowledgments}
National Natural Science Foundation of China (11805141); Fundamental Research Program of Shanxi Province (202303021211185); Israel Science Foundation (grant No. 1695/22); ANID (Chile) through FONDECYT (1260401).
\end{acknowledgments}


\begin{thebibliography}{99}
\bibitem{Ref1} C. Elphick and E. Meron, \textquotedblleft Localized
structures in surface waves,\textquotedblright\ Phys. Rev. A \textbf{40},
3226 (1989).

\bibitem{Ref2} W. Zhang and J. Vi\~nals, \textquotedblleft Secondary
instabilities and spatiotemporal chaos in parametric surface
waves,\textquotedblright\ Phys. Rev. Lett. \textbf{74}, 690 (1995).

\bibitem{Ref3} X. Wang and R. Wei, \textquotedblleft Oscillatory patterns
composed of the parametrically excited surface-wave
solitons,\textquotedblright\ Phys. Rev. E \textbf{57}, 2405 (1998).

\bibitem{Ref4} J. N. Kutz, W. L. Kath, R. D. Li, and P. Kumar,
\textquotedblleft Long-distance pulse propagation in nonlinear optical
fibers by using periodically spaced parametric
amplifiers,\textquotedblright\ Opt. Lett. \textbf{18}, 802--804 (1993).

\bibitem{Ref5} I. V. Barashenkov, M. M. Bogdan, and V. I. Korobov,
\textquotedblleft Stability diagram of the phase-locked solitons in the
parametrically driven, damped nonlinear Schr\"{o}dinger
equation,\textquotedblright\ Europhys. Lett. \textbf{15}, 113 (1991).

\bibitem{Ref6} A. Mecozzi, W. L. Kath, P. Kumar, and C. G. Goedde,
\textquotedblleft Long-term storage of a soliton bit stream by use of
phase-sensitive amplification,\textquotedblright\ Opt. Lett. \textbf{19},
2050--2052 (1994).

\bibitem{Ref7} S. Longhi, \textquotedblleft Ultrashort-pulse generation in
degenerate optical parametric oscillators,\textquotedblright\ Opt. Lett.
\textbf{20}, 695--697 (1995).

\bibitem{Ref8} S. Longhi, \textquotedblleft Stable multipulse states in a
nonlinear dispersive cavity with parametric gain,\textquotedblright\ Phys.
Rev. E \textbf{53}, 5520 (1996).

\bibitem{Ref9} P. Grelu, and N. Akhmediev, \textquotedblleft Dissipative
solitons for mode-locked lasers,\textquotedblright\ Nat. Photonics \textbf{6}%
, 84--92 (2012).

\bibitem{Ref10} N. Englebert, F. D. Lucia, P. Parra-Rivas, C. M. Arab\'{\i},
P.-J. Sazio, S.-P. Gorza, and F. Leo, \textquotedblleft Parametrically
driven Kerr cavity solitons,\textquotedblright\ Nat. Photonics 15, 857--861
(2021).

\bibitem{Ref11} G. Moille, M. Leonhardt, D. Paligora, N. Englebert, F. Leo,
J. Fatome, K. Srinivasan, and M. Erkintalo, \textquotedblleft Parametrically
driven pure-Kerr temporal solitons in a chip-integrated
microcavity,\textquotedblright\ Nat. Photon. 18, 617--624 (2024).

\bibitem{Ref12} Y. Geng, H. Zhou, X. Han, W. Cui, Q. Zhang, B. Liu, G. Deng,
Q. Zhou, and K. Qiu, \textquotedblleft Coherent optical communications using
coherence cloned Kerr soliton microcombs,\textquotedblright\ Nat. Commun.
13, 1070 (2022).

\bibitem{Ref13} S. A. Diddams, K. Vahala, and T. Udem, \textquotedblleft
Optical frequency combs: Coherently uniting the electromagnetic
spectrum,\textquotedblright\ Science. 369, 6501, (2020).

\bibitem{Ref14} H. Shu, L. Chang, Y. Tao, B. Shen, W. Xie, M. Jin, A.
Netherton, Z. Tao, X. Zhang, R. Chen, B. Bai, J. Qin, S. Yu, X. Wang, and J.
E. Bowers, \textquotedblleft Microcomb-driven silicon photonic
systems,\textquotedblright\ Nature 605, 457 (2022).

\bibitem{Ref15} A. L. Gaeta, M. Lipson, and T. J. Kippenberg,
\textquotedblleft Photonic-chip-based frequency combs,\textquotedblright\
Nat. Photonics 13, 158 (2019).

\bibitem{Ref16} I. V. Barashenkov and E. V. Zemlyanaya, \textquotedblleft
Stable complexes of parametrically driven, damped nonlinear Schr\"{o}dinger
solitons,\textquotedblright\ Phys. Rev. Lett. \textbf{83}, 2568 (1999).

\bibitem{Ref17} N. V. Alexeeva, I. V. Barashenkov, and D. E. Pelinovsky,
\textquotedblleft Dynamics of the parametrically driven NLS solitons beyond
the onset of the oscillatory instability,\textquotedblright\ Nonlinearity
\textbf{12}, 103 (1999).

\bibitem{Ref18} I. V. Barashenkov, S. R. Woodford, and E. V. Zemlyanaya,
\textquotedblleft Parametrically driven dark solitons,\textquotedblright\
Phys. Rev. Lett. \textbf{90}, 054103 (2003).

\bibitem{Ref19} I. V. Barashenkov, E. V. Zemlyanaya, and M. B\"{a}r,
\textquotedblleft Traveling solitons in the parametrically driven nonlinear
Schr\"{o}dinger equation,\textquotedblright\ Phys. Rev. E \textbf{64},
016603 (2001).

\bibitem{Ref19a} I. V. Barashenkov and E. V. Zemlyanaya,
\textquotedblleft Traveling solitons in the externally driven nonlinear
Schr\"{o}dinger equation,\textquotedblright\ J. Phys. A: Math. Theor. \textbf{44}, 465211 (2011).

\bibitem{Ref20} I. V. Barashenkov and E. V. Zemlyanaya, \textquotedblleft
Traveling solitons in the damped-driven nonlinear Schr\"{o}dinger
equation,\textquotedblright\ SIAM J. Appl. Math. \textbf{64}, 800--818
(2004).

\bibitem{Ref21} A. O. Le\'{o}n, M. G. Clerc, and S. Coulibaly,
\textquotedblleft Traveling pulse on a periodic background in parametrically
driven systems,\textquotedblright\ Phys. Rev. E \textbf{91}, 050901 (2015).

\bibitem{Ref22} I. V. Barashenkov and E. V. Zemlyanaya, \textquotedblleft
Soliton complexity in the damped-driven nonlinear Schr\"{o}dinger equation:
Stationary to periodic to quasiperiodic complexes,\textquotedblright\ Phys.
Rev. E \textbf{83}, 056610 (2011).

\bibitem{Ref23} I. V. Barashenkov, S. R. Woodford, and E. V. Zemlyanaya,
\textquotedblleft Interactions of parametrically driven dark solitons. I. N%
\'{e}el--N\'{e}el and Bloch--Bloch interactions,\textquotedblright\ Phys.
Rev. E \textbf{75}, 026604 (2007).

\bibitem{Ref24} I. V. Barashenkov and S. R. Woodford, \textquotedblleft
Interactions of parametrically driven dark solitons. II. N\'{e}el--Bloch
interactions,\textquotedblright\ Phys. Rev. E \textbf{75}, 026605 (2007).

\bibitem{new}
Y. Zhang. W. Feng, Q. Wang, X. Ren, X. Zhang, D. Zou, and Z. Li,
\textquotedblleft Optical solitons dominated by pure-high-even-order dispersion: Research progress of pure-quartic solitons,\textquotedblright\ Laser Photonics Rev. \textbf{2026}, e71177 (2026).

\bibitem{Ref25} A. Blanco-Redondo, C. M. de Sterke, J. E. Sipe, T. F Krauss,
B. J. Eggleton, and C. Husko, \textquotedblleft Pure-quartic
solitons,\textquotedblright\ Nat. Commun. \textbf{7}, 10427 (2016).

\bibitem{Ref26} A. F. J. Runge, D. D. Hudson, K. K. K. Tam, C. M. de Sterke, and A. Blanco-Redondo, \textquotedblleft The pure-quartic soliton laser,\textquotedblright\ Nat. Photonics \textbf{14}, 492--497 (2020).

\bibitem{Ref27} Z. Wang, Y. Lin, Y. Song, Z. Deng, J. Liu, and C. Zhang,
\textquotedblleft Revealing the dynamics of multiple solitons and soliton
pulsations in a quartic-dispersion fiber laser,\textquotedblright\ Laser
Photonics Rev. \textbf{20}, e01851 (2026).

\bibitem{Ref28} Z. C. Qian, M. Liu, A. P. Luo, Z. C. Luo, and W. C. Xu,
\textquotedblleft Dissipative pure-quartic soliton fiber
laser,\textquotedblright\ Opt. Express \textbf{30}, 22066--22073 (2022).

\bibitem{Ref29} Z. L. Wu, M. Liu, Y. X. Gao, Z. X. Zhang, M. Luo, Y. Hu, T.
J. Li, A. P. Luo, W. C. Xu, and Z. C. Luo, \textquotedblleft Spectral
sidebands of dissipative soliton in a positive fourth-order-dispersion fiber
laser,\textquotedblright\ Opt. Express \textbf{32}, 47882--47892 (2024).

\bibitem{Ref30} Z. L. Wu, Z. X. Zhang, T. J. Li, M. Luo, M. Liu, A. P. Luo,
W. C. Xu, and Z. C. Luo, \textquotedblleft Pulsating dynamics of dissipative
pure-quartic soliton in a fiber laser,\textquotedblright\ Opt. Express
\textbf{33}, 10129--10139 (2025).

\bibitem{Ref31} H. Taheri and A. B. Matsko, \textquotedblleft Quartic
dissipative solitons in optical Kerr cavities,\textquotedblright\ Opt. Lett.
\textbf{44}, 3086--3089 (2019).

\bibitem{Ref32} P. Parra-Rivas, S. Hetzel, Y. V. Kartashov, P. F. de C\'{o}%
rdoba, J. A. Conejero, A. Aceves,\ and C. Mili\'{a}n, \textquotedblleft
Quartic Kerr cavity combs: bright and dark solitons,\textquotedblright\ Opt.
Lett. \textbf{47}, 2438-2441 (2022).

\bibitem{Ref33} K. Liu, S. Yao, and C. Yang, \textquotedblleft Raman pure
quartic solitons in Kerr microresonators,\textquotedblright\ Opt. Lett.
\textbf{46}, 993-996 (2021).

\bibitem{Ref34} S. Mei, M. Liu, H. Huang, T. He, Z. Lu, and W. Zhao,
\textquotedblleft Route to pure-quartic solitons in the
mid-infrared,\textquotedblright\ Opt. Lett. \textbf{50}, 2848--2851 (2025).

\bibitem{Ref35} H. Zhu, M. Liu, H. Huang, J. Wang, S. Mei, and Wei Zhao,
\textquotedblleft Raman-assisted pure quartic platicons in
microresonators,\textquotedblright\ Opt. Lett. \textbf{50}, 3357--3360
(2025).

\bibitem{Ref36} C. Silvestri, Y. Qiang, K. Panda, J. Widjaja, S. Coen, C. M.
de Sterke, and A. F. J. Runge, \textquotedblleft Pure high-order dispersion
dissipative Kerr solitons in optical cavities,\textquotedblright\ Opt. Lett.%
\textbf{50}, 4262--4265 (2025).

\bibitem{Ref37} J. Yang, \textit{Nonlinear waves in integrable and
nonintegrable systems }(SIAM, Philadelphia, 2010).
\end{thebibliography}
\end{document}